\documentclass[acmsmall]{acmart}
\usepackage{etex}

\usepackage{subcaption} 

\acmConference[]{}

\usepackage{booktabs}   
\usepackage{subcaption} 
\usepackage[noend]{algpseudocode}

\author{Yuan Si}
\affiliation{%
  \institution{University of Waterloo}
  \city{Waterloo}
  \country{Canada}
  }
\email{yuan.si@uwaterloo.ca}

\author{Simeng Han}
\affiliation{%
  \institution{Stanford University}
  \city{Stanford}
  \country{USA}
  }
\email{shan6@law.stanford.edu}

\author{Daming Li}
\affiliation{%
  \institution{Independent Researcher}
  \country{USA}
  }
\email{damingliyale22@gmail.com}

\author{Hanyuan Shi}
\affiliation{%
  \institution{Independent Researcher}
  \country{China}
  }
\email{shihanyuan1995@gmail.com}

\author{Jialu Zhang}
\authornote{Corresponding Author}
\affiliation{%
  \institution{University of Waterloo}
  \city{Waterloo}
  \country{Canada}
  }
\email{jialu.zhang@uwaterloo.ca}

\usepackage{graphicx}
\usepackage{amsmath}

\let\emptyset\varnothing
\usepackage{xspace}
\usepackage{footnote}
\usepackage{amsfonts}
\usepackage{natbib}
\usepackage{hhline}
\usepackage{tabularx}
\usepackage{scratch3}
\usepackage{diagbox}
\usepackage{multirow}
\usepackage{setspace}
\usepackage{epsfig}
\usepackage{url}

\usepackage{breakurl}
\usepackage{booktabs}
\usepackage{tabularx}
\newcolumntype{Y}{>{\raggedright\arraybackslash}X}
\usepackage{tikz}
\usetikzlibrary{arrows.meta,positioning,shapes.geometric,shapes.misc,fit,backgrounds}

\usepackage{xcolor}
\usepackage{lipsum}
\usepackage{courier}
\usepackage{listings}
\usepackage{caption}
\usepackage{colortbl}
\usepackage{arydshln} 
\usepackage{makecell} 
\usepackage{tikz}
\usepackage{color}
\usepackage{rotating}

\usepackage{varwidth}
\usetikzlibrary{fit,arrows,calc,positioning,quotes}

\usepackage{pgfplots}

\usepackage{mathtools}
\usepackage{fancyvrb}
\usepackage{textcomp}

\usepackage{longtable}
\usepackage{pdflscape} 
\usepackage{threeparttable} 
\usepackage{array}     

\usepackage{stmaryrd}

\usepackage{algorithm}
\usepackage[noend]{algpseudocode}

\newcolumntype{P}[1]{>{\raggedright\arraybackslash}p{#1}}

\usepackage{alltt}

\tikzset{
  -|-/.style={
    to path={
      (\tikztostart) -| ($(\tikztostart)!#1!(\tikztotarget)$) |- (\tikztotarget)
      \tikztonodes
    }
  },
  -|-/.default=0.5,
  |-|/.style={
    to path={
      (\tikztostart) |- ($(\tikztostart)!#1!(\tikztotarget)$) -| (\tikztotarget)
      \tikztonodes
    }
  },
  |-|/.default=0.5
}

\usetikzlibrary{shapes.geometric}

\long\def\com#1{}

\newcommand{\app}{\texttt{ScratchEval}\xspace}

\newcommand{\para}[1]{\smallskip\noindent {\bf #1}}

\newcommand{\squishlist}{
   \begin{list}{$\bullet$}
    { \setlength{\itemsep}{0pt}      \setlength{\parsep}{3pt}
      \setlength{\topsep}{3pt}       \setlength{\partopsep}{0pt}
      \setlength{\leftmargin}{3.5mm} \setlength{\labelwidth}{1em}
      \setlength{\labelsep}{0.5em} }
}

\newcommand{\squishend}{
    \end{list}  }

\def\BibTeX{{\rm B\kern-.05em{\sc i\kern-.025em b}\kern-.08em
    T\kern-.1667em\lower.7ex\hbox{E}\kern-.125emX}}

\begin{document}

\title{\app: A Multimodal Evaluation Framework for LLMs in Block-Based Programming}

\begin{abstract}
	LLMs have achieved strong performance on text-based programming tasks, yet they remain unreliable for block-based languages such as Scratch. Scratch programs exhibit deeply nested, non-linear structures, event-driven concurrency across multiple sprites, and tight coupling between code and multimedia assets, properties that differ fundamentally from textual code. As a result, LLMs often misinterpret Scratch semantics and generate large, invasive edits that are syntactically valid but semantically incorrect when repairing buggy programs.

We introduce ScratchEval, the first executable benchmark designed to evaluate LLM-based repair for Scratch programs, covering program understanding, debugging, analysis, and repair. The benchmark contains 100 curated Scratch projects from the public repository, selected for structural and semantic complexity. Each project is paired with executable test suites, bug descriptions with corresponding fixes, block-level edit constraints defining minimal semantically correct repairs, and required multimedia assets. The benchmark is constructed through a human-in-the-loop pipeline combining automated project mining with expert validation of trigger-outcome semantics and representative bug patterns, with emphasis on event ordering, concurrency, and state management.

To enable rigorous and reproducible evaluation, we propose a three-layer executable protocol measuring functional correctness via VM-level execution, repair quality using block-level edit distance and behavioral trajectory comparisons, and explanation quality via structured rubrics assessing alignment between model reasoning and generated patches. Using ScratchEval, we study domain-specific fine-tuning, training data effectiveness, and model generalization to unseen bug types. ScratchEval provides a reproducible foundation for evaluating and post-training LLMs on block-based programming tasks.
\end{abstract}

\maketitle

\section{Introduction}
\label{sec:intro}
LLMs have recently demonstrated strong capabilities across a wide range of programming tasks, including code generation, debugging, and automated repair for text-based languages such as Python, Java, and C/C++. These successes are driven in part by large-scale pre-training on textual code corpora and by benchmarks that enable fine-grained, executable evaluation of functional correctness and repair quality. In contrast, much less is understood about LLM performance on block-based programming languages, which differ fundamentally from textual languages in program structure, execution model, and representation.

Scratch is a widely used block-based programming environment with tens of millions of shared projects\footnote{\url{https://scratch.mit.edu/statistics/}}. Programs consist of nested blocks distributed across sprites and coordinated by broadcasts and implicit concurrency, so behavior depends on cross-script interactions, shared state, and tightly coupled media assets~\cite{resnick2009scratch, maloney2010scratch,verifiedscratch20}. 
This execution model challenges current LLMs: they often misinterpret triggers, event ordering, and synchronization, producing syntactically valid but semantically off-target and frequently over-invasive repairs~\cite{verifiedscratch20,QiLAR15,whisker_modelbased2022}.

Despite the importance of Scratch and similar educational programming environments, progress in this space has been hindered by the lack of rigorous, executable benchmarks tailored to block-based languages. Existing evaluations often rely on small, ad hoc examples and manual inspection of model outputs, or use coarse pass/fail criteria that fail to assess repair quality, semantic fidelity, or explanation correctness~\cite{SiStitch2025}. Although large numbers of Scratch projects can be scraped, such corpora are dominated by trivial or low-information samples and lack ground-truth bug descriptions, gold-standard fixes, and executable test oracles~\cite{fraser2021litterbox,moreno2015drscratch}. Consequently, it remains difficult to systematically evaluate LLM capabilities on Scratch, to compare models fairly, or to study the effects of domain-specific training in this domain.

We argue that LLM support for block-based programming needs a small but execution-rigorous benchmark. \app provides semantically dense Scratch projects that expose representative mechanisms (event-driven concurrency, state management, and cross-sprite interactions) and enable precise, VM-executable evaluation across understanding, debugging, and repair.

The benchmark consists of 100 Scratch projects drawn from the public Scratch repository, each selected for high structural and semantic complexity. Every project is paired with (1) an executable test suite that runs in a Scratch virtual machine, (2) a detailed bug description and gold-standard fix, (3) block-level edit constraints defining minimal, semantically faithful repairs, and (4) the multimedia assets required for faithful execution. We construct the benchmark via a human-in-the-loop pipeline that combines automated project mining with expert validation of trigger–mechanism–outcome semantics and representative bug patterns, with particular focus on concurrency, event ordering, and shared state.

We introduce a three-layer, executable evaluation protocol that separates functional correctness, repair quality, and explanation faithfulness under Scratch’s event-driven execution model. First, \textbf{functional correctness} is assessed via automated execution in a Scratch virtual machine: the model’s patch is applied to the original project and tested using injected events and assertions. Second, \textbf{repair quality} is measured in terms of patch minimality and semantic fidelity, using block-level edit distance and behavioral trajectory comparisons against the gold fix. Third, \textbf{explanation quality} is evaluated using a structured rubric to determine whether the model’s explanation correctly and consistently aligns with its proposed repair. 
This protocol goes beyond a binary pass/fail outcome by checking that a fix is small, targeted, and preserves the intended behavior.

Using \app, we investigate several fundamental research questions, including the impact of domain-specific fine-tuning on repair success and quality, the effectiveness of different training data sources, and the ability of models to generalize to unseen types of bugs. Our study provides new insights into the limitations of current LLMs on block-based programming and establishes a closed-loop framework for evaluating and improving LLMs in this domain.

\para{Why ISSTA?} \app targets the core concerns of the ISSTA community, including executable testing, automated repair, and oracle design, in a programming paradigm where these problems are fundamentally harder due to event-driven concurrency and multimodal state. Unlike prior Scratch datasets or educational tools,  \app formalizes what it means to test and repair block-based programs under rigorous, VM-level execution, making it a natural extension of ISSTA’s long-standing work on Defects4J-style benchmarks and test-driven repair.

This paper makes the following contributions:
\begin{itemize}
    \item \textbf{Evaluation Methodology.} We introduce a VM-executable, multi-layer evaluation methodology for block-based programs that measures functional correctness, repair minimality, semantic fidelity, and explanation grounding under Scratch’s event-driven execution model.
    \item \textbf{Benchmark.} We present \app, a high-density benchmark of 100 Scratch projects with controlled, testable bugs, minimal gold repairs, and executable test oracles.
    \item \textbf{Empirical Study.} We conduct a systematic empirical evaluation of state-of-the-art LLMs, analyzing their program understanding, bug diagnosis, and repair behavior.
\end{itemize}

\section{Background and Motivating Examples}
\label{sec:motiv}

\begin{figure}
    \centering
    \includegraphics[width=\linewidth]{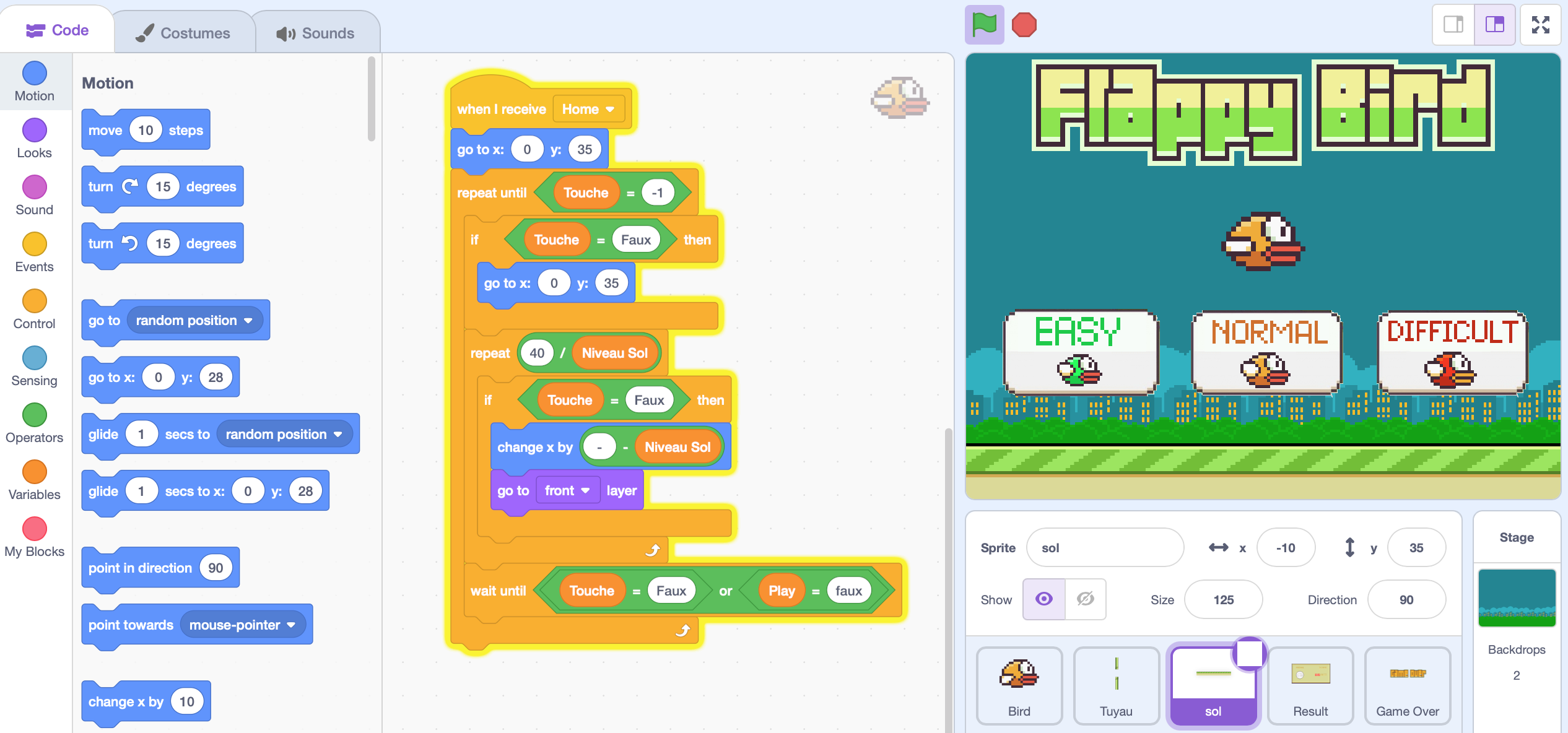}
    \caption{An example Scratch project of "Flappy Bird" game. Source:~\url{https://scratch.mit.edu/projects/195385320}}
    \label{fig:gui}
\end{figure}

\para{Background.} Scratch is a block-based visual programming environment in which programs are assembled as stacks of connected blocks and executed by sprites on a stage. As illustrated in Fig.~\ref{fig:gui}, the interface comprises four primary areas: the block palette (left) containing categorized commands, the scripting workspace (center) for composing logic, the stage (upper right) for visualizing execution, and the sprite pane (lower right) for managing independent actors and their assets. Each script responds to events (e.g., key presses or broadcasts), and program behavior emerges from interactions among independent, asynchronous, event-driven scripts. Scratch further couples code with multimodal assets such as graphics, costumes, and sounds, which directly affect program behavior. 

The features including nested visual blocks, event-driven concurrency, shared state, and multimedia integration, make Scratch program analysis fundamentally different from that of text-based languages~\cite{resnick2009scratch,maloney2010scratch}.

Despite Scratch’s widespread use, progress in automated support for block-based programs has been limited by the lack of rigorous, executable benchmarks. Prior evaluations have relied on small ad hoc examples or manual inspection, without ground-truth fixes or reliable test oracles. As a result, the community has lacked a systematic way to evaluate LLMs on Scratch programs, motivating our work on \app.

\para{Motivating Examples.}
To illustrate the challenges, consider a few typical bug scenarios in Scratch. 

\para{Example 1 (Concurrency).}
A game synchronizes two sprites using a \texttt{broadcast-and-wait} block; replacing it with a plain \texttt{broadcast} introduces a race condition in which one sprite proceeds before the other is ready. The correct repair is a single local change that restores the missing synchronization. In contrast, LLM-generated patches often modify or reorder unrelated blocks, failing to preserve the intended concurrency semantics.

\para{Example 2 (State initialization).}
A score variable should be reset to zero at the start of each level, but a missing initialization causes values to accumulate across runs. The minimal fix is to insert or relocate a single initialization block. LLMs, however, often introduce additional logic or unrelated edits, producing repairs that are structurally larger than necessary.

\para{Example 3 (Clone management).}
Scratch programs frequently create sprite clones that must be explicitly initialized or deleted. Failing to do so leads to clone accumulation and incorrect behavior, typically fixable by adding a single \texttt{delete this clone} block. LLM-generated repairs often affect other scripts or shared state, resulting in unnecessarily invasive patches.

\para{Example 4 (Visibility toggle).}
A sprite's visibility is toggled via a key press using a boolean flag and Scratch’s built-in \texttt{show}/\texttt{hide} blocks. When this logic is faulty, the minimal repair is a small, localized change to the toggle condition. LLMs may instead introduce auxiliary counters or arithmetic state, increasing complexity and introducing new failure modes despite the availability of a simpler boolean solution.

In fact, users on Scratch's official forums\footnote{\url{https://scratch.mit.edu/discuss/}} have observed that LLM-generated code “tend[s] to make up blocks, [produce] code that’s impossible, overly complex, or just nonfunctional,” which makes such fixes hard to adopt and debug\footnote{\url{https://scratch.mit.edu/discuss/topic/824561/?page=3}}. These examples illustrate common bug patterns in Scratch (e.g., missing initialization, synchronization errors, clone leaks) and underscore the need for an evaluation approach that emphasizes precise, minimal repairs.

\section{System Design}
\label{sec:design}

In this section, we describe \app's end-to-end pipeline for constructing an executable Scratch benchmark and evaluating LLMs under a VM-level execution contract (Fig.~\ref{figs:system-framework}). Given a public Scratch project as input, our pipeline produces a self-contained benchmark instance with five key components: (i) a runnable \texttt{.sb3} project snapshot along with all its assets (images and sounds); (ii) a buggy variant of the project created by one controlled, reversible block-level edit; (iii) the inverse of that edit, representing the minimal correct fix; (iv) an executable test suite (a set of interaction scenarios with rerun-stabilized oracle assertions) to validate the project’s behavior in the Scratch VM; and (v) a reference semantic record documenting the project’s triggers, mechanisms, and expected outcomes for later comprehension evaluation. We rely only on code and assets shipped with each project (no external gameplay videos) and implement the pipeline in five stages: project collection, expert curation, bug synthesis/annotation, test-oracle synthesis, and evaluation integration (Sections~\ref{sec:design:Project Selection and Curation}--\ref{sec:design:Evaluation Protocol Integration}).

\app is designed around two goals: (1) faithfully exercising Scratch-specific difficulties (event-driven concurrency, non-linear control flow, and code--asset coupling), and (2) enabling automated, precise, and reproducible evaluation via VM-level execution and structured artifacts~\cite{verifiedscratch20, stahlbauer2019whisker, Si2025ViScratch,SiStitch2025}.

\subsection{Architecture Overview}
\label{sec:design:Architecture Overview}
\begin{figure}
    \centering
    \includegraphics[width=\linewidth]{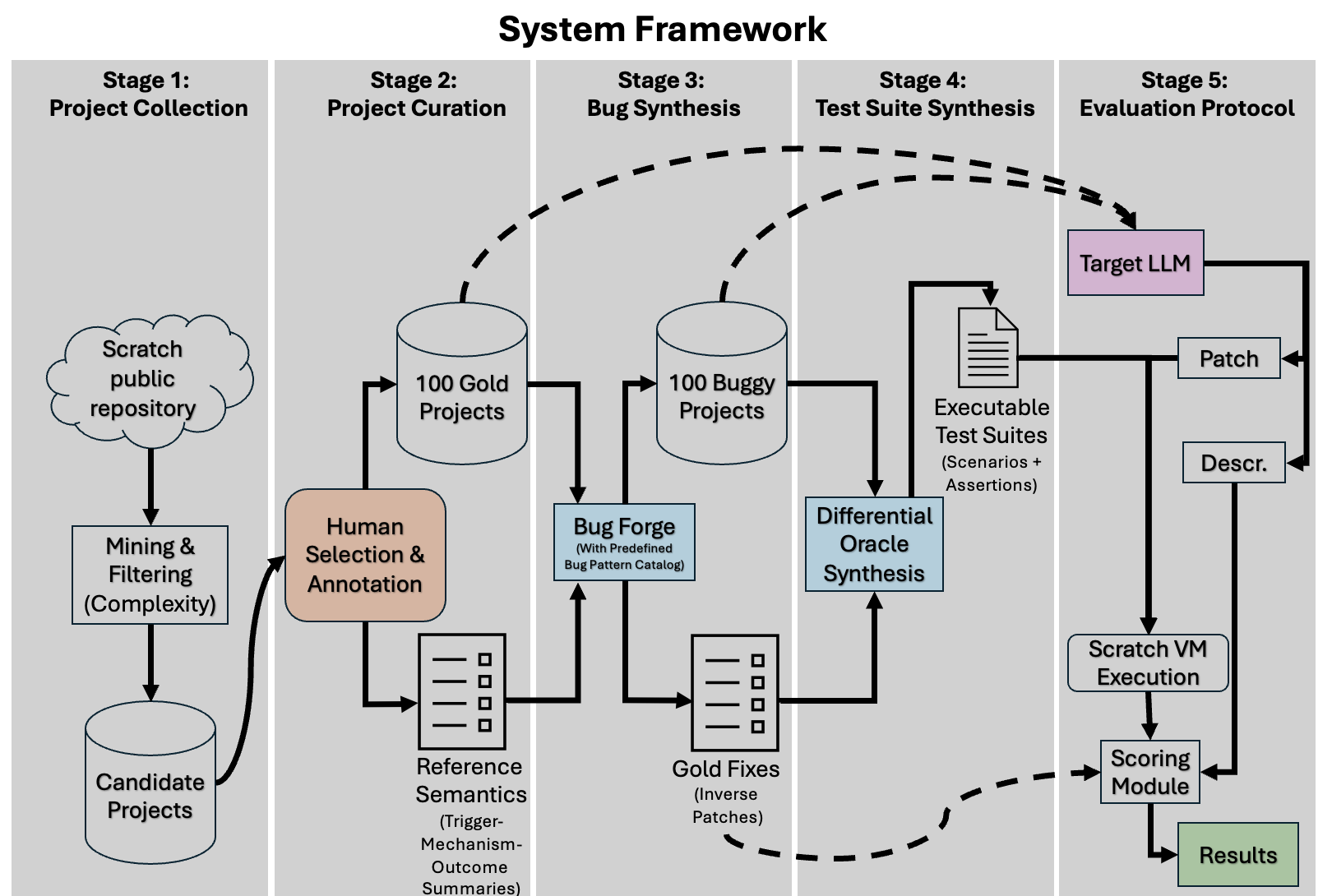}
    \caption{System framework of \app. Stages~1--2 mine and filter the Scratch repository, then curate 100 gold projects with reference semantics (trigger--mechanism--outcome summaries and key state signals). Stages~3--4 inject one reversible bug per project (Bug Forge) and synthesize executable tests via rerun-stabilized differential oracles from tick-level traces. Stage~5 applies LLM patches, executes projects in the Scratch VM, and scores functional correctness and repair/understanding quality.}
    \label{figs:system-framework}
\end{figure}

Fig.~\ref{figs:system-framework} summarizes the end-to-end pipeline. \app is artifact-driven: each stage consumes and produces machine-readable artifacts that form the only interface between stages, enabling deterministic replay given the same inputs and tool versions. This modular design also supports ablations and extensions (e.g., new bug operators or oracle checks) without changing the model-facing interface.

 \subsection{Project Selection and Curation}
 \label{sec:design:Project Selection and Curation}
\para{Project Selection.}
We crawl the public Scratch repository to download \texttt{.sb3} projects and filter for
interaction-rich programs (multiple sprites/scripts, broadcasts, custom blocks, clones, and asset use).
We retain projects meeting fixed complexity thresholds (e.g., $\ge5$ sprites, $\ge15$ scripts,
$\ge3$ broadcasts, and at least one custom block), yielding a few hundred candidates.

\para{Expert Curation.}
Two authors manually review the candidates and select 100 semantically non-trivial projects spanning
diverse genres (games/animations/stories), excluding near-duplicates and ensuring broad feature
coverage (timing/sensing, concurrency, lists, clones, and broadcasts).

\para{Complexity Criteria.}
We filter projects using fixed thresholds: at least $N{=}5$ sprites, $M{=}15$ scripts, $E{=}3$ broadcast uses, and at least one custom block definition, which reduces tens of thousands of projects to a few hundred high-complexity candidates.

\para{Human Expert Inductive Curation.} Next, we performed a manual curation step on the filtered candidates. Two authors independently reviewed the remaining projects to select the final set of 100 for \app. During this curation, we considered not only the quantitative metrics above but also each project’s semantic complexity and uniqueness. We favored projects that implement meaningful behaviors (e.g., a game with scoring and levels, an interactive story with user choices, a physics-based simulation) and that had clear points of potential failure or difficulty where an LLM might struggle.

To ensure diversity, we excluded near-duplicate projects. The final set spans multiple domains (games, animations, stories) and exercises a broad range of Scratch features (e.g., timing/sensing inputs, parallel scripts, lists, clones, and broadcasts). 

\para{Reference Semantics Record.} For each selected gold project, we create a lightweight, machine-readable reference semantics record that anchors later evaluation. The record captures (i) a project-level overview (goal, primary sprites/roles, and the dominant interaction loop), (ii) a catalog of key event hooks that a user can trigger (e.g., green-flag start, key presses, sprite clicks, and broadcasts) together with the involved sprites, and (iii) the intended outcomes of those hooks expressed in VM-observable signals. We additionally record a compact set of \emph{state signals} (e.g., variables/lists, sprite pose/costume/visibility, backdrop, and broadcasts) that our harness can log deterministically and that downstream tools can reuse.

\para{Separation from Repair Ground Truth.} This record is designed to describe \emph{what} the project is intended to do and \emph{what} should be observable when key interactions occur, without encoding fix-specific hints. In particular, it does \emph{not} expose the injected edit location or the minimal gold repair; those remain ground truth for scoring and test synthesis (Sections~\ref{sec:design:Bug Synthesis and Annotation}--\ref{sec:design:Test Suite Generation and Oracles}). This separation prevents accidental leakage while still enabling consistent scoring of global understanding and trigger/outcome grounding.

\subsection{Bug Synthesis and Annotation}
\label{sec:design:Bug Synthesis and Annotation}
\para{Identifying and Injecting Bugs.}
For each curated project, Bug Forge injects exactly one fault by applying a single reversible, block-level transformation over the Scratch JSON IR, producing a buggy variant that remains parsable and executable by the Scratch VM.
We restrict injections to a catalog of semantics-aware operators that (i) induce an observable behavioral deviation under interaction, (ii) admit an unambiguous inverse transformation that we treat as the minimal gold repair, and (iii) localize the edit to a small, well-defined region of the block graph to support minimality analysis.

Each injected instance is recorded as structured, machine-readable metadata: the operator type, a textual description of expected behavior, and the precise injection site in the IR, together with paired forward/inverse patch representations. These records guide scenario-and-oracle construction and provide ground truth for judging whether model-generated explanations correctly identify the intended trigger, mechanism, and outcome~\cite{Si2025ViScratch}.

\para{Bug Pattern Catalog and Automation.}
We curated a catalog of \textbf{eight representative bug patterns} to cover a broad range of Scratch defects. To ground this catalog in real evidence, we began with ViScratch’s bug list and incorporated additional bug types reported in prior Scratch analyses and LLM debugging studies~\citep{Si2025ViScratch,stahlbauer2019whisker,verifiedscratch20,fraser2021litterbox,fein2025litterboxplus,SiStitch2025}. We merged overlapping categories and kept only patterns meeting three criteria: (i) they can be injected as small, localized edits in the Scratch JSON; (ii) they have a well-defined inverse fix (reversible edit); and (iii) they produce a reliably observable behavior difference under our automated test harness. Table~\ref{tab:bugpatterns} shows the patterns with their injection operators and the failure modes they covered. The patterns include Missing Initialization; Desynchronization (Missing Wait); Untriggered Event (Message Omitted/Mismatched); Non-Terminating Loop/Condition; Incorrect Conditional Logic; Sprite State Mismatch; Clone Management Error; and Event Handler Conflict (Stutter).

\newcolumntype{P}[1]{>{\raggedright\arraybackslash}p{#1}}

\begin{table}[t]
\centering
\small
\begin{tabularx}{\linewidth}{@{}P{0.23\linewidth} Y Y@{}}
\toprule
\textbf{Pattern} & \textbf{Injection operator (typical edit)} & \textbf{Covered failure modes (examples)} \\
\midrule
Missing Initialization
& Delete/move/reset a variable initialization (e.g., remove \texttt{set score to 0} or perturb its literal)
& State leakage across rounds; score/lives never reset; stale mode flags \\ \addlinespace

Desynchronization (Missing Wait)
& \texttt{broadcast and wait} $\rightarrow$ \texttt{broadcast}, or remove a short \texttt{wait}
& Race conditions; message order violations; read-before-ready bugs \\ \addlinespace

Untriggered Event\\(Message Omitted/Mismatched)
& Mutate message name or remove an emission/receiver binding
& Handlers never fire; missing game-over/win transitions; UI never updates \\ \addlinespace

Non-Terminating Loop/Condition
& Flip termination condition or perturb threshold literal
& Infinite play; level never ends; win/lose unreachable \\ \addlinespace

Incorrect Conditional Logic
& Flip comparator/opcode or swap operands/variables
& Off-by-one; wrong branching; inverted win/collision logic \\ \addlinespace

Sprite State Mismatch
& Remove/move costume/visibility updates or swap a target costume/backdrop
& Wrong visual state; invisible sprite; incorrect animation/facing \\ \addlinespace

Clone Management Error
& Perturb clone lifecycle (miss deletion / missing per-clone init)
& Clone accumulation; inconsistent clone state; uncontrolled growth \\ \addlinespace

Event Handler Conflict (Stutter)
& Duplicate/overlap handlers or remove gating checks
& Double-move, stutter, repeated transitions from unsynchronized handlers \\
\bottomrule
\end{tabularx}
\caption{Bug patterns and covered failure modes in \app.}
\label{tab:bugpatterns}
\end{table}

\para{Bug Forge: Controlled and Testable Fault Injection.}
We implement these bug patterns in an automated engine, Bug Forge, which applies a chosen pattern to a gold project’s JSON IR and emits a localized forward patch plus its inverse, enabling consistent and reproducible bug planting. Bug Forge is designed with two key principles in mind:

\begin{itemize}
    \item \textbf{Controlled Minimality:} Each pattern is realized as a small, localized edit to the project’s block-level JSON IR (e.g., mutating a single block opcode/field or removing one initialization block), ensuring the injected fault is isolated.
    \item \textbf{Testability:} An injected bug is accepted only if it reliably induces an externally observable failure under the synthesized scenarios—i.e., the gold project passes while the buggy variant fails at least one oracle assertion across reruns~\cite{jia2011mutation}.
\end{itemize}

To enforce the testability criterion, \textsc{Bug Forge} operates an inject-and-validate loop. For a given project and bug pattern, the engine will: select a candidate location in the project to inject the bug; apply the pattern’s code transformation; then run the project with our test generation pipeline to check outcomes. It verifies that the original “gold” project passes all tests while the newly injected “buggy” project fails at least one oracle assertion. If the buggy project does not produce a detectable failure (i.e., tests still pass), the injection is discarded or adjusted, and the engine may try a different injection point. This loop repeats until the bug reliably triggers a failure, guaranteeing that each accepted bug in \app is coupled with a strong test oracle. Pseudocode for this process is provided in Algorithm~\ref{alg:bugforge}. Unless otherwise specified, we set $R{=}5$, $\theta_{\text{pass}}{=}0.9$, $\theta_{\text{fail}}{=}0.1$, and $K{=}20$ in this inject-and-validate loop. 

\begin{algorithm}[t]
\caption{\textsc{Bug Forge}: inject-and-validate loop}
\label{alg:bugforge}
\begin{algorithmic}[1]
\Require Gold project $P^{\text{gold}}$, pattern catalog $\mathcal{B}$ (8 patterns),
template library $\mathcal{T}$, reruns $R$, thresholds $(\theta_{\text{pass}}, \theta_{\text{fail}})$, max trials $K$
\Ensure Buggy project $P^{\text{bug}}$, forward patch $\Delta_{\text{gold}\rightarrow\text{bug}}$,
inverse patch $\Delta_{\text{bug}\rightarrow\text{gold}}$, bug spec $S_{\text{bug}}$, testsuite $T$
\State $b \gets \Call{SelectPattern}{\mathcal{B}}$ \Comment{coverage-aware scheduling}
\For{$i \gets 1$ to $K$}
  \State $loc \gets \Call{SelectInjectionSite}{P^{\text{gold}}, b}$ \Comment{prefer T--M--O chain}
  \State $(P', \Delta) \gets \Call{ApplyOperator}{P^{\text{gold}}, b, loc}$ \Comment{localized reversible edit}
  \State $\mathcal{M} \gets \Call{ExtractMetadata}{P^{\text{gold}}}$ \Comment{keys/messages/state signals}
  \State $\mathcal{S} \gets \Call{ScenarioInstantiate}{P^{\text{gold}}, \mathcal{T}, \mathcal{M}}$ \Comment{Alg.~\ref{alg:scenario-inst}}
  \State $T \gets \Call{SynthesizeTestSuite}{P^{\text{gold}}, P', \mathcal{S}, R, \theta_{\text{pass}}, \theta_{\text{fail}}}$ \Comment{Alg.~\ref{alg:testsynth}}
  \State $p_g \gets \Call{PassRate}{P^{\text{gold}}, T, R}$ \Comment{multi-run robustness}
  \State $p_b \gets \Call{PassRate}{P', T, R}$
  \If{$p_g \ge \theta_{\text{pass}}$ \textbf{and} $p_b \le \theta_{\text{fail}}$}
    \State $P^{\text{bug}} \gets P'$
    \State $\Delta_{\text{gold}\rightarrow\text{bug}} \gets \Delta$
    \State $\Delta_{\text{bug}\rightarrow\text{gold}} \gets \Call{InversePatch}{\Delta}$
    \State $S_{\text{bug}} \gets \Call{EmitBugSpec}{b, loc, \Delta, T}$
    \State \Return $(P^{\text{bug}}, \Delta_{\text{gold}\rightarrow\text{bug}}, \Delta_{\text{bug}\rightarrow\text{gold}}, S_{\text{bug}}, T)$
  \EndIf
\EndFor
\State \Return \textbf{fail}
\end{algorithmic}
\end{algorithm}

Algorithm~\ref{alg:bugforge} formalizes this automation, reducing manual effort and ensuring consistent bug injection. Each bug pattern is implemented as a small transformation function (e.g., removing a variable initialization under the green flag script for Missing Initialization, or replacing broadcast-and-wait with a non-waiting broadcast for Desynchronization); after injection, we verify that the project still loads and runs in the Scratch VM (albeit exhibiting the buggy behavior).

Importantly, the distribution of these patterns across the 100 projects is roughly balanced: each bug pattern is represented by multiple projects (each pattern appears in at least about 10 projects, and at most around 17), ensuring that no single pattern dominates the benchmark. This diversity compels models to handle different types of bugs rather than overfit to one particular bug type.

\para{Structured Bug Record (Model-facing vs. Ground Truth).} Alongside each injected buggy variant, \app emits a structured bug record. The record contains a \emph{model-facing bug report} (observed symptom, expected correct behavior, and a reproducible trigger interaction), as well as \emph{ground-truth metadata} used only by our pipeline: the injected mechanism label, the injection site (sprite/script/block identifiers in the JSON IR), and the inverse patch as the minimal gold fix. We enforce that the inverse patch is minimal and semantically faithful: it resolves the injected fault without altering unrelated behavior.

The ground-truth fields drive test-oracle synthesis and automatic scoring, but they are \emph{not} revealed to evaluated models. Models must infer the trigger--mechanism--outcome explanation and/or localize and repair the fault from the buggy project and the bug report alone.

\subsection{Test Suite Generation and Oracles}
\label{sec:design:Test Suite Generation and Oracles}
\para{Automated Test Harness.}
We execute projects in the official Scratch~3.0 VM\footnote{\url{https://github.com/scratchfoundation/scratch-editor/tree/develop/packages/scratch-vm}} via its public API, programmatically replaying interaction scenarios (green-flag, key presses, sprite clicks, broadcasts) while stepping execution tick-by-tick. The harness logs VM-observable state (variables/lists, sprite/backdrop/costume/visibility,
and broadcasts) at checkpoints and evaluates assertions automatically, producing traces used for differential oracle synthesis.

\begin{algorithm}[t]
\caption{Scenario instantiation from templates}
\label{alg:scenario-inst}
\begin{algorithmic}[1]
\Require Gold project $P^{\text{gold}}$, template library $\mathcal{T}$, metadata $\mathcal{M}$ (triggers/messages/keys/clickable sprites)
\Ensure Instantiated scenario set $\mathcal{S}$
\If{$\mathcal{M} = \emptyset$}
  \State $\mathcal{M} \gets \Call{ExtractMetadata}{P^{\text{gold}}}$ \Comment{fallback if metadata is not provided}
\EndIf
\State $\mathcal{S} \gets \emptyset$
\State $\mathcal{K} \gets \Call{ExtractKeys}{\mathcal{M}}$ \Comment{e.g., \texttt{space}, \texttt{left arrow}}
\State $\mathcal{C} \gets \Call{ExtractClickableSprites}{\mathcal{M}}$
\State $\mathcal{B} \gets \Call{ExtractBroadcastMessages}{\mathcal{M}}$
\ForAll{$T \in \mathcal{T}$}
  \State $params \gets \Call{SelectParams}{T, \mathcal{K}, \mathcal{C}, \mathcal{B}}$
  \State $s \gets \Call{Instantiate}{T, params}$ \Comment{fill concrete keys/sprites/messages}
  \State $\mathcal{S} \gets \mathcal{S} \cup \{s\}$
\EndFor
\State $\mathcal{S} \gets \Call{Deduplicate}{\mathcal{S}}$
\State \Return $\mathcal{S}$
\end{algorithmic}
\end{algorithm}

\para{Deriving Test Cases from Bugs.}
We derive focused test cases from each bug record (trigger and expected outcome). Specifically, we instantiate interaction templates with project metadata (keys, clickable sprites, and broadcast messages; Algorithm~2) and assert the expected state at a small set of checkpoints.
In most cases, a single short interaction sequence is sufficient; for example, for a ``game does not end when lives reach 0'' bug, the scenario drives the life counter to zero and checks for the corresponding game-over signal (e.g., a broadcast or backdrop change).

\para{Handling Randomness.} One challenge in testing Scratch projects is their frequent use of randomness. For instance, a game might position objects at random coordinates (using Scratch’s \texttt{pick random} blocks), which can lead to inconsistent outcomes across test runs. If left uncontrolled, such non-determinism could make our tests flaky and the logs cluttered with irrelevant variations. To address this, we fix the random seed for the Scratch RNG (random number generator) within our test harness for each scenario, so that RNG-driven behaviors are consistent across runs, following established practice in Scratch testing~\citep{stahlbauer2019whisker, deiner2023autotest}. 

\para{Rerun-stabilized Oracle Filtering.} Even with a fixed RNG seed, residual non-determinism can arise from event timing and concurrent scheduling. We therefore execute each synthesized scenario for $R$ reruns when selecting oracle assertions. Unless noted, we use $R{=}5$, a tick budget of $H{=}2000$, and checkpoints every 10 ticks (plus the final tick).

\begin{algorithm}[t]
\caption{Differential testsuite synthesis}
\label{alg:testsynth}
\begin{algorithmic}[1]
\Require $P^{\text{gold}}$, $P^{\text{bug}}$, scenario set $\mathcal{S}$,
reruns $R$, tick budget $H$, checkpoints $\mathcal{C}$,
thresholds $(\theta_{\text{pass}}, \theta_{\text{fail}})$
\Ensure Testsuite $T=(\mathcal{S}, \mathcal{A})$
\State $\mathcal{A} \gets \emptyset$
\ForAll{$s \in \mathcal{S}$}
  \For{$r \gets 1$ to $R$}
    \State \Call{SetRandomSeed}{$r$} \Comment{or fixed seed per scenario}
    \State $\tau^{g}_{r} \gets \Call{RunVMByTick}{P^{\text{gold}}, s, H, \mathcal{C}}$
    \State \Call{SetRandomSeed}{$r$}
    \State $\tau^{b}_{r} \gets \Call{RunVMByTick}{P^{\text{bug}}, s, H, \mathcal{C}}$
  \EndFor
  \State $\mathcal{F} \gets \Call{ExtractFeatures}{\{\tau^{g}_{r}\}_{r=1}^{R}, \{\tau^{b}_{r}\}_{r=1}^{R}}$
  \ForAll{$f \in \mathcal{F}$}
    \State $p_g \gets \Call{HoldProb}{f, \{\tau^{g}_{r}\}_{r=1}^{R}}$
    \State $p_b \gets \Call{HoldProb}{f, \{\tau^{b}_{r}\}_{r=1}^{R}}$
    \If{$p_g \ge \theta_{\text{pass}}$ \textbf{and} $p_b \le \theta_{\text{fail}}$}
      \State $\mathcal{A} \gets \mathcal{A} \cup \{\Call{MakeAssertion}{f}\}$
    \EndIf
  \EndFor
\EndFor
\State \Return $T=(\mathcal{S}, \mathcal{A})$
\end{algorithmic}
\end{algorithm}

\para{Differential Oracle Synthesis.} 
Our oracle assertions are synthesized by comparing execution traces between the gold and buggy versions. Specifically, for each test scenario, we run the gold project and buggy project in parallel (under the same inputs) and record certain key state variables over time. We then identify state signals that reliably differ between the correct and buggy runs. For example, a gold project might set variable “score” to 10 after a certain event, while the buggy version does not, or the gold might broadcast “win” while the buggy does not. Such differences can form the basis of an oracle: if a candidate fix is correct, its execution trace should match the gold project on these signals; if not, it might resemble the buggy trace.

To operationalize this stability, in Algorithm~\ref{alg:testsynth}, we calculate the probability $p_g$ that the signal holds in the gold run and $p_b$ that it holds in the buggy run. We accept an assertion only if $p_g$ is high (above $\theta_{\text{pass}}$) and $p_b$ is low (below $\theta_{\text{fail}}$). The resulting test suite includes scenario scripts plus a set of assertions that check program state at key points. In practice, each project’s test suite might contain a handful of assertions (e.g., 3--5) covering different aspects of the intended behavior~\cite{mckeeman1998differential}.

\subsection{Evaluation Protocol Integration}
\label{sec:design:Evaluation Protocol Integration}
Finally, \app integrates all components into a unified, model-facing evaluation protocol. Each benchmark instance defines an explicit \emph{model contract}: (i) the buggy Scratch project in a standardized representation derived from Scratch~3.0's JSON IR (preserving stable identifiers for sprites/variables/blocks so edits can be referenced precisely), and (ii) a short bug report describing the observed symptom and the expected behavior under a reproducible interaction. Crucially, we do \emph{not} provide the injection site, the gold fix, or any execution traces to the model.

Depending on the track, the model produces either (a) a structured patch that can be applied to the JSON IR (repair) or (b) a structured explanation of project and bug semantics (understanding). \app then scores outputs automatically: for repair, it applies the patch and executes the patched project in the Scratch VM under synthesized tests; for understanding, it compares the structured fields against the reference semantics and bug records.

\para{Repair Evaluation.} In the repair track, an LLM outputs a patch in a structured, machine-applicable format. Concretely, a patch is a JSON list of \emph{atomic block edits} over the Scratch JSON IR, where each edit is one of: (i) \texttt{remove} a block (with well-defined reconnection of its neighbors), (ii) \texttt{add} a new block with an explicit parent/next linkage, or (iii) \texttt{modify} an existing block by replacing a specific opcode/field/input. Each edit identifies its target by \texttt{sprite\_id} and \texttt{block\_id} (and, when needed, a field/input path), enabling unambiguous application. We treat schema-invalid or non-applicable patches as failures.

We apply the model patch to the buggy project, execute the patched project in the Scratch VM under the synthesized test suite, and count a repair as \emph{functionally successful} iff the project loads, does not crash, and passes all assertions.

Beyond success, we quantify patch quality. Let $E_{\text{gold}}$ be the set of normalized atomic edits in the minimal gold fix (inverse patch) and $E_{\text{model}}$ the set for the model patch. We define \emph{edit distance} as the symmetric-difference size
$ d_{\text{edit}} = |E_{\text{gold}} \triangle E_{\text{model}}| $,
which counts both missing gold edits and unnecessary extra edits. We define \emph{semantic drift} as the average normalized trace distance between the patched project and the gold project on the same scenarios: we log VM-observable state signals (e.g., variables/lists, sprite pose/costume/visibility, backdrop, broadcasts) at fixed checkpoints and compute the mean per-signal discrepancy across checkpoints and scenarios (0 = identical to gold, 1 = maximally different). Together, $d_{\text{edit}}$ and drift capture whether a patch is minimal and behavior-preserving rather than a plausible but invasive rewrite.

\para{Understanding Evaluation.} In the understanding track, the LLM outputs a structured explanation with two parts: (i) a \emph{global project summary} (goal, core sprites/roles, and the main interaction loop), and (ii) a \emph{bug explanation} in trigger--mechanism--outcome (TMO) form. The trigger describes the event/interaction that activates the faulty behavior; the mechanism is a categorical label from our bug taxonomy; and the outcome describes the observed deviation from the intended behavior. The model may additionally provide an optional free-form narrative.

We score the global summary against the Stage~2 reference semantics record to obtain global understanding accuracy (G-Acc). For bug understanding, we compare the TMO fields against ground truth from the bug record: triggers and outcomes are matched after canonicalizing event descriptions into a fixed vocabulary (e.g., \texttt{green\_flag}, \texttt{key:space}, \texttt{broadcast:game\_over}); when canonicalization is ambiguous, we fall back to a conservative fuzzy match. Mechanisms are scored by exact match on the mechanism tag. We report strict joint correctness (U-Acc; correct iff all three TMO fields are correct), trigger identification F1, and mechanism accuracy. Finally, we score each free-form explanation with a rubric-guided LLM judge (Table~\ref{tab:rubric}) to quantify explanation quality beyond the structured fields.

\para{Reproducibility and Robustness.} We enforce strict reproducibility throughout evaluation: VM executions are deterministic given a fixed seed, and oracle selection is rerun-stabilized (Section~\ref{sec:design:Test Suite Generation and Oracles}).

\section{Evaluation}
\label{sec:evaluation}
We evaluate three representative LLMs (\textsc{ChatGPT~5.2}\footnote{\url{https://platform.openai.com/docs/models/gpt-5.2}}, \textsc{Gemini~3}\footnote{\url{https://ai.google.dev/gemini-api/docs/models}}, and \textsc{Qwen~3}\footnote{\url{https://www.alibabacloud.com/help/en/model-studio/what-is-qwen-llm}}) on \app’s repair and understanding tasks using the unified evaluation protocol defined in Section~\ref{sec:design:Evaluation Protocol Integration}. All models are run zero-shot using greedy (deterministic) decoding (i.e., no sampling; temperature $T=0$).
 For the open-weight \textsc{Qwen} model, we additionally report a lightweight LoRA adaptation under a two-fold held-out protocol (50/50 split), while closed models are evaluated zero-shot only~\cite{hu2021lora}.

\para{Inputs.} Each instance provides the buggy project exported from \texttt{.sb3} (blocks, sprites, and metadata) together with the project’s original assets (PNG/SVG images and WAV sounds). We do not provide gameplay videos.

\para{Inference Configuration.} All zero-shot runs use 
decoding (temperature $=0$, $\texttt{top\_p}=1.0$) with a 2048-token output cap; each instance is attempted once ($n=1$).

\para{Execution Environment.}
Unless otherwise noted, all VM execution, oracle checking, and metric computation were run on a Mac mini (Apple M4, 10-core CPU, 16\,GB memory) with macOS~26.2, Node.js~v20.19.1, and Python~3.9.6.
Our harness pins \texttt{scratch-vm} v5.0.300 to avoid behavioral drift across VM versions.
For closed models, inference runs on the providers' infrastructure; all outputs are evaluated locally using the same VM-level pipeline.

\para{LoRA Setup (Qwen only).}
We fine-tune LoRA adapters while keeping the Qwen~3 base checkpoint frozen.
We use two-fold project-level cross evaluation (disjoint halves $A/B$, 50 each): tune on $A$ and evaluate on held-out $B$, and vice versa, so each project is evaluated exactly once without leakage.
Unless otherwise noted, we use rank $r{=}16$, scaling $\alpha{=}32$, dropout 0.05, learning rate $2\times10^{-4}$, 3 epochs, batch size 4, and max sequence length 4096. Optimization uses AdamW with PyTorch default hyperparameters~\cite{loshchilov2019decoupled,pytorchAdamW}.

\para{Human Calibration for LLM-as-a-judge.}
To check the reliability of our rubric-guided LLM judge, we sample 20 instances (stratified by bug mechanism) and have two authors independently score model outputs using the same rubric and structured schema.
We measure inter-annotator agreement using Cohen's $\kappa$ (weighted when applicable) and adjudicate disagreements by discussion.
The judge agrees with the adjudicated labels on 85\% of cases, with substantial chance-corrected agreement ($\kappa{=}0.74$, 95\% CI [0.50, 0.90]).

We answer the following RQs:
\begin{itemize}
    \item \textbf{RQ1 (G-Acc):} Can models recover project-level intent and structure, and how does this relate to bug-level understanding?
    \item \textbf{RQ2 (U-Acc):} How accurately do models explain a bug’s trigger--mechanism--outcome?
    \item \textbf{RQ3 (T-F1):} How well do models identify bug triggers?
    \item \textbf{RQ4 (M-Acc):} How accurately do models classify bug mechanisms?
    \item \textbf{RQ5 (Repair):} How often do models produce a functionally correct fix, and how minimal/faithful are their patches?
\end{itemize}
Unless stated otherwise, we report results on all 100 projects; LoRA results apply to \textsc{Qwen} under the two-fold held-out protocol.

We next present results for each RQ with quantitative metrics and analysis. \textbf{Table~\ref{tab:metrics}} summarizes core understanding metrics in the zero-shot setting, and (for \textsc{Qwen}) the corresponding LoRA-tuned results and repair gains.

\begin{table}[t]
\centering
\small
\setlength{\tabcolsep}{4pt}

\begin{tabular}{llccccc}
\toprule
\textbf{Model} & \textbf{Setting} & \textbf{Global (G-Acc)} & \textbf{Bug (U-Acc)} & \textbf{Trigger F1} & \textbf{Mechanism Acc} & \textbf{Fix Gains} \\
\midrule
\textsc{ChatGPT} & Zero-shot  & 79\% & 57\% & 0.52 & 0.62 & -- \\
\textsc{Gemini}  & Zero-shot  & 81\% & 66\% & 0.55 & 0.65 & -- \\
\textsc{Qwen}    & Zero-shot  & 59\% & 40\% & 0.38 & 0.50 & -- \\
\textsc{Qwen}    & LoRA-tuned & 65\% & 42\% & 0.45 & 0.58 & +3 \\
\bottomrule
\end{tabular}
\caption{Zero-shot and LoRA-tuned performance on \app evaluation metrics. Closed models (\textsc{ChatGPT} and \textsc{Gemini}) are evaluated zero-shot only; the LoRA-tuned setting applies to \textsc{Qwen} (open weights). 
Global Understanding (G-Acc) = accuracy of recovering the project-level intent/structure;
Bug Understanding (U-Acc) = percentage of bugs where the model’s structured explanation matches the reference (trigger, mechanism, and outcome);
T-F1 = trigger identification F1-score; M-Acc = mechanism tag accuracy;
Fix Gains = number of previously failing \emph{repairs} that became correct after LoRA tuning (i.e., fail$\rightarrow$success relative to \textsc{Qwen} zero-shot on the 100-bug evaluation set).}
\label{tab:metrics}
\end{table}

\subsection{RQ1: Global Project Understanding (G-Acc)}
We mainly measure whether a model can recover the \textbf{global intent} and \textbf{functional structure} of a Scratch project (e.g., the project goal, main loop, core sprites, and major interactions).
This is distinct from bug-level understanding: a model may describe what the project is supposed to do while still failing to ground \emph{why} the buggy behavior occurs.

In the baseline setting, LLMs already demonstrated strong global understanding:
\textsc{Gemini} reaches 81\% accuracy, slightly higher than \textsc{ChatGPT} (79\%), while \textsc{Qwen} trails at 59\%.
This observation is consistent with our earlier analysis that the main bottleneck is often not ``what the project is about'' but ``why the bug happens'' under Scratch's event-driven concurrency.

For \textsc{Qwen}, LoRA tuning improves global understanding from 59\% to 65\% (Table~\ref{tab:metrics}). We assess statistical significance with a two-sided exact McNemar test on paired per-project correctness outcomes; the discordant counts are $n_{01}=6$ (incorrect$\rightarrow$correct) and $n_{10}=0$ (correct$\rightarrow$incorrect), yielding $p=3.13\times10^{-2}$ ($N=100$)~\cite{mcnemar1947note}.

\subsection{RQ2: Bug Understanding Accuracy}
\label{secs:bugUnderstanding}
This metric evaluates the LLMs’ structured comprehension of each bug, i.e., whether the model correctly identifies the bug’s trigger, mechanism, and outcome in its explanation.
U-Acc is a strict joint metric: an instance is counted correct only if trigger, mechanism, and outcome are all correct; RQ3 and RQ4 then break down performance into trigger-only and mechanism-only components.
Table~\ref{tab:metrics} shows that all models exhibit limited \textbf{baseline understanding} of Scratch bugs: \textsc{ChatGPT} correctly understood 57\% of the bugs, \textsc{Gemini} 66\%, and \textsc{Qwen} 40\%. These values reflect frequent mismatches between model explanations and the true event-driven causes.

For \textsc{Qwen}, LoRA tuning yields a small increase in understanding accuracy (40\% $\to$ 42\%; Table~\ref{tab:metrics}), suggesting modest gains in structured bug understanding from limited in-domain data. Since \textsc{ChatGPT} and \textsc{Gemini} are evaluated zero-shot only, we do not report post-adaptation U-Acc for them. Overall, structured bug understanding remains challenging, especially for event-driven concurrency scenarios.

\para{Rubric-based Explanation Scoring.} In addition to strict U-Acc, we evaluate the model’s free-form bug explanation using a rubric-guided LLM judge (Section~\ref{sec:design:Evaluation Protocol Integration}). The judge assigns a 1--5 score based on the following criteria:

\begin{table}[t]
\centering
\small
\begin{tabular}{p{0.2\linewidth}p{0.75\linewidth}}
\toprule
\textbf{Score} & \textbf{Criteria (Bug Explanation Rubric)} \\
\midrule
1 (Poor) & Explanation is mostly incorrect or irrelevant; fails to identify trigger or outcome; provides no meaningful insight. \\
2 (Weak) & Identifies some correct symptom, but misattributes cause; trigger/mechanism mostly wrong or vague. \\
3 (Fair) & Correctly identifies trigger or outcome, but incomplete or partially incorrect mechanism; some grounding in Scratch elements. \\
4 (Good) & Correctly identifies trigger, mechanism, and outcome with minor omissions; grounded in specific blocks/events. \\
5 (Excellent) & Fully correct and detailed explanation; clearly links trigger$\rightarrow$mechanism$\rightarrow$outcome with strong grounding in project details. \\
\bottomrule
\end{tabular}
\caption{Rubric for scoring free-form bug explanations (1--5).}
\label{tab:rubric}
\end{table}

In the zero-shot setting, the closed models (\textsc{ChatGPT} and \textsc{Gemini}) generally produce more grounded and consistent narratives than \textsc{Qwen}, and are more likely to reference concrete Scratch elements (sprite names, broadcast messages, variables) rather than speaking in generalities. For \textsc{Qwen}, LoRA tuning yields a modest improvement in grounding and consistency, but a substantial fraction of explanations still deviate from the true trigger or mechanism (Table~\ref{tab:metrics}). In summary, \textbf{limited in-domain adaptation helps, but event-driven bug understanding remains challenging}.

\subsection{RQ3: Trigger Identification (T-F1)}
Next, we assess the models’ ability to identify the \textbf{trigger} of each bug, defined as the specific event or action that initiates the faulty behavior, measured using the F1-score against the gold-standard triggers. As shown in Table~\ref{tab:metrics}, baseline trigger identification was poor (\textsc{Qwen}: 0.38; \textsc{ChatGPT}: 0.52; \textsc{Gemini}: 0.55), indicating that even strong models miss many true triggers.

For \textsc{Qwen}, LoRA tuning improves trigger detection from 0.38 to 0.45 (Table~\ref{tab:metrics}). Closed models are evaluated zero-shot only, so we do not report post-adaptation trigger F1 for \textsc{ChatGPT} and \textsc{Gemini}. Despite the gain on \textsc{Qwen}, trigger identification remains challenging: even the best zero-shot model reaches only 0.55 F1, reflecting persistent difficulty with event interdependencies.

\subsection{RQ4: Mechanism Classification (M-Acc)}
We define mechanism accuracy as how often the model correctly identifies or classifies the underlying bug mechanism (the causal category of the bug), such as concurrency/timing, state initialization, or missing handlers. Baseline mechanism tagging was moderate (\textsc{ChatGPT}: 62\%; \textsc{Gemini}: 65\%; \textsc{Qwen}: 50\%; Table~\ref{tab:metrics}), reflecting the challenge of mapping event-driven behavior to causal categories.

For \textsc{Qwen}, LoRA tuning increases mechanism classification accuracy from 50\% to 58\%
(Table~\ref{tab:metrics}).
Closed models are evaluated in the zero-shot setting only; accordingly, we do not report
post-adaptation M-Acc for \textsc{ChatGPT} or \textsc{Gemini}.
Despite this improvement, mechanism diagnosis remains challenging:
across models, approximately 35--50\% of instances are still misclassified,
with errors concentrated in concurrency-intensive scenarios.

\subsection{RQ5: Repair Success and Quality}
Finally, we evaluate the models on the ultimate task: fixing the bug. We measure \textbf{functional success rate} (the percentage of bugs fixed so all tests pass) and also examine patch quality (minimality and semantic preservation).
For the \textsc{Qwen} LoRA-tuned setting, we follow the same two-fold $A/B$ protocol described above: each buggy project is evaluated only in the fold where it is held out from tuning, and we aggregate the two held-out halves to obtain $N=100$ LoRA-tuned outcomes.

Even under zero-shot conditions, \textbf{correct repairs are rare}. \textsc{Gemini} fixed 41\% of the bugs, \textsc{ChatGPT} 32\%, and \textsc{Qwen} only 23\%. Moreover, the fixes that models do produce are often not minimal. On average, a model’s patch differs from the gold fix by about 4 block edits, whereas the intended gold fixes typically require just a single-block change. We also observed a significant degree of semantic drift: unintended side-effects in behavior. The average drift score was around 0.30 (on a 0–1 scale), meaning that even when a patch passes all the tests, it tends to alter other behaviors compared to the original intended execution.

After applying LoRA fine-tuning, \textsc{Qwen}’s success rate rose slightly from 23\% to 26\% (fixing 3 additional bugs). A McNemar’s test on the paired outcomes shows this increase is not statistically significant ($p \approx 0.25$, with 3 fail$\rightarrow$success improvements and 0 success$\rightarrow$fail regressions out of 100 cases)~\cite{mcnemar1947note}. However, LoRA had a clear positive effect on patch quality. The average edit distance fell from about 4 blocks down to ~2 blocks, and the drift score dropped from 0.30 to 0.15. This indicates that the fine-tuned model’s patches were much closer to minimal edits and preserved the projects’ behavior more faithfully.

We categorize common repair failures into three recurring modes: (1) \textbf{over-editing}, where models introduce auxiliary state/logic beyond the localized fix; (2) \textbf{incomplete fixes}, where a patch addresses a symptom but misses required updates elsewhere (e.g., reset one state variable but not a coupled flag); and (3) \textbf{mis-grounded concurrency fixes}, where models apply sequential-code heuristics that do not respect Scratch’s event ordering (e.g., replacing synchronization with counters/loops). Overall, zero-shot repair success is low (23\%--41\%), and even when fixes succeed they frequently deviate from minimality; lightweight in-domain adaptation yields modest success gains but substantial improvements in patch cleanliness. These results highlight event-driven Scratch debugging as a challenging frontier for LLM repair and motivate future work on stronger execution-grounded reasoning and Scratch-specific semantic modeling.

\section{Discussion}
\label{sec:discussion}

\app's evaluation reveals a recurring failure pattern in LLM-based repair for Scratch:
models often produce patches that are locally plausible at the level of an individual
script, yet fail under VM execution. We refer to these as \emph{near-miss patches}.
In our benchmark, near-misses rarely arise from malformed or ill-typed edits.
Instead, they violate \emph{whole-program semantic commitments} imposed by
Scratch’s event-driven runtime and its serialized \texttt{.sb3} representation.
This section distills two representative classes of near-miss failures and discusses
their implications for repair systems targeting block-based, concurrent programs.

\subsection{Near-miss Patches Violate Semantic, not Syntactic, Constraints}
Across projects, near-miss patches concentrate on two categories of constraints that
are difficult to preserve when reasoning locally.
First, \textbf{temporal constraints} arise from Scratch’s synchronization primitives,
such as the barrier semantics of \texttt{broadcast and wait}.
Second, \textbf{binding constraints} arise from the \texttt{.sb3} representation, in which
blocks reference variables and sprites via stable identifiers rather than surface labels.
Both constraints are enforced by the VM during execution and are explicitly encoded
in \texttt{.sb3} metadata, making them amenable to automatic, lightweight checking.
These failures therefore reflect semantic mismatches rather than  syntactic
errors.

\subsection{Case Study: Restoring Synchronization Barriers}
In \app project~1197499245, the injected fault replaces the
\texttt{broadcast and wait} with the \texttt{broadcast}. The message name stays the same, but the meaning changes:
\texttt{broadcast and wait} pauses the sender until every script that handles the message finishes, acting as a synchronization
barrier~\cite{ScratchBroadcasting,ScratchVMIssue2109,ScratchWikiBroadcastAndWait}. Without the wait, the sender can continue
immediately, so later blocks may run before receivers update shared state.
Many near-miss patches try to ``fake'' the barrier with a short \texttt{wait} or by polling a flag variable.
But a fixed delay cannot predict how long receivers will take, and polling changes the schedule by adding extra work.
Such patches may pass simple runs, yet they fail under different event schedules.
For this class of bugs, repairing should be \emph{barrier-aware}: if the oracle expects receivers to finish before the next
phase, the patch space should prefer restoring \texttt{broadcast and wait} itself rather than heuristic timing substitutes.

\subsection{Case Study: Repairing \texttt{.sb3} Bindings, not Just Labels}
In \app project~1197377838, a \texttt{set variable} block updates \texttt{target direction}
when it should update \texttt{x speed}. The gold repair is a single-field change that
redirects the block to the correct variable.
The subtlety is that \texttt{.sb3} is not name-based.
Variables are represented by stable identifiers in the project JSON, and blocks reference
those identifiers in their fields~\cite{DeepWikiSB3Format,ScratchVMIssue1372}.
As a result, modifying only the displayed variable name is insufficient.
Two variables may appear identical at the UI level while having distinct IDs
(e.g., a global variable versus a sprite-local one).
A near-miss patch may therefore rename a variable or select a similarly named variable
from the wrong scope; the script appears correct, yet the VM continues to read from or
write to the original identifier at runtime.
For such defects, repair must be \emph{binding-aware}:
the intended \texttt{.sb3}-level reference—both identifier and scope—must be updated
consistently across all reads and writes, rather than altering only the surface label.

\subsection{Size of the Dataset: Why Selecting 100 Scratch Projects?}
\app prioritizes \emph{semantic density and executable rigor} over raw scale. Unlike large scraped corpora dominated by trivial or non-interactive projects~\cite{Aivaloglou2016HowKidsCode,Aivaloglou2017ScratchDataset}, each \app instance is manually curated, paired with VM-executable tests, and annotated with reversible, minimal repairs. As a result, each project supports multiple orthogonal evaluations (understanding, diagnosis, repair correctness, minimality, and explanation grounding), making \app substantially denser than large but weakly-labeled corpora. This design mirrors Defects4J-style benchmarks, where depth and executability are favored over breadth.

\subsection{Implications for Repair Systems}
These observations suggest a practical design principle for LLM-based repair of
block-structured, event-driven programs: patch generation should be paired with
lightweight \emph{semantic validators} that encode the runtime’s coordination and
binding contracts. Concretely, repair systems can (i) treat synchronization primitives
as first-class edit targets and penalize heuristic delay substitutions when a barrier
is required~\cite{ScratchVMIssue2109,ScratchWikiBroadcastAndWait}, and (ii) verify
that candidate edits modify \texttt{.sb3}-level references (IDs and scopes) consistently
across all uses~\cite{DeepWikiSB3Format,ScratchVMIssue1372}. These checks are local,
machine-checkable, and complementary to VM-level testing: they filter or re-rank
plausible-but-invalid patches before execution, reducing wasted trials and discouraging
over-editing.

Near-miss patches serve as a useful diagnostic signal, revealing where local plausibility
diverges from executable correctness. By making these gaps measurable under a VM
oracle and minimality constraints, \app motivates repair pipelines that explicitly
enforce temporal and binding semantics, rather than relying solely on text- or
block-level heuristics.
\section{Threats to Validity}
\label{validity}

\para{Internal Validity.}
Our results rely on the correctness of the evaluation pipeline (parsing, patch application, and VM execution). To reduce pipeline errors, we enforce a machine-parsable patch format and reject ambiguous outputs. We validate the pipeline by running the harness on buggy projects and gold fixes, confirming that the former fail and the latter pass. Scratch projects may exhibit nondeterminism due to timing and event interleavings. Although we fix random seeds and use rerun-stabilized oracle selection (Section~\ref{sec:design:Test Suite Generation and Oracles}), residual variability can remain, and a small number of borderline cases may persist.

\para{Construct Validity.}
Functional correctness is measured by whether a candidate repair passes the
synthesized, VM-executed test suites.
While execution-based oracles are objective and reproducible, they are necessarily
incomplete: passing a finite test suite does not guarantee semantic correctness,
and test-passing patches may still be wrong for untested behaviors.
This limitation is well known in generate-and-validate repair.
We mitigate this by deriving oracles from gold and buggy differential traces under controlled reruns and by augmenting pass or fail outcomes with minimality and behavioral drift metrics, though regressions outside our scenarios may still remain.
Repair-quality metrics (e.g., block-level edit distance) approximate minimality and semantic fidelity but do not
capture all subjective notions of patch acceptability.
Explanation quality is evaluated using a rubric-guided LLM judge; although we fix
the rubric, configuration, and judge model, some preference or calibration bias may
persist~\cite{liu2023geval}.

\para{External Validity.}
\app currently contains 100 curated projects, each with a single bug. We intentionally limit each instance to one error to enable controlled evaluation of bug understanding and repair; however, this design does not capture scenarios with multiple interacting bugs.
Bugs in \app are injected using predefined patterns. Although these patterns are derived from issues commonly observed in real Scratch projects, they may not cover all naturally occurring bugs in the wild. Also, some Scratch projects or code patterns may have appeared in the pre-training data of the evaluated models, particularly closed-source models. This possibility is difficult to rule out and may affect the interpretation of generalization results.

\para{Conclusion Validity.}
We evaluate a fixed set of LLMs and prompts under a fixed decoding configuration. Different models, prompting strategies (e.g., tool use or iterative refinement), or decoding may yield different results. Moreover, given the benchmark size, small differences in aggregate accuracy may not be statistically robust. Accordingly, we avoid over-interpreting marginal deltas and focus on consistent trends across metrics, bug categories, and evaluation dimensions.

Despite these limitations, we prioritize rigor and reproducibility through curated instances, controlled bug injection, and executable, VM-level evaluation. While \app does not cover all possible Scratch scenarios, it provides an auditable foundation for studying LLM behavior on block-based program understanding and repair. We view \app as a starting point that can be extended in future work with additional projects, bug patterns, and evaluation refinements.

\section{Related Work}
\label{relat}

\para{Code and Repair Benchmarks.} Prior work on ML for code has produced several benchmarks, but almost all target text-based languages. For example, OpenAI’s HumanEval is a well-known Python code generation benchmark (164 handwritten tasks with test suites) used to measure LLM's code generation capability~\cite{chen2021codex}. Microsoft’s CodeXGLUE suite offers a dozen datasets for various code understanding and generation tasks~\cite{codexglue2021}. SWE-bench provides a benchmark of GitHub issues for evaluating LLMs on end-to-end software engineering tasks~\cite{jimenez2024swebench}. In program repair, Defects4J is a classic dataset of real Java bugs (hundreds of faults with accompanying test suites) designed for reproducible evaluation~\cite{JustJE2014Defects4J}. However, none of these include graphical or block-based programs. Recent studies (e.g. Ibrahimzada et al.) have also explored synthetic bug generation for ML-based repair (e.g. BUGFARM~\cite{Ibrahimzada2025SyntheticBugs}), but again in textual code. Our \app fills a gap by providing a block-based counterpart: a curated, executable benchmark specifically for Scratch.

\para{Scratch and Block-based Programming.} Several projects have addressed analysis or AI support for Scratch. Stahlbauer et al. developed Bastet, a framework for formal analysis of Scratch (using an intermediate language)~\cite{verifiedscratch20}, and also a testing framework for Scratch programs. These tools target program verification and testing, not LLM-based repair. Complementary to analysis and testing, Deiner and Fraser introduced NuzzleBug, an omniscient and interrogative debugger for Scratch that supports breakpoints, stepping (including reverse stepping), and question answering over executions to help learners understand program behavior~\cite{deiner2024nuzzlebug}. More directly, Schweikl and Fraser proposed RePurr, an automated repair approach for Scratch learners' programs based on evolutionary search~\cite{schweikl2025repurr}. In AI-assisted Scratch, ViScratch uses both the project’s code and a generated gameplay video as inputs to diagnose and fix bugs~\cite{Si2025ViScratch}. This multimodal approach significantly improves repair quality, highlighting the importance of visual context. Similarly, Scratch Copilot provides an AI tutoring assistant for children in a Scratch-like environment~\cite{DrugaOtero2023}, helping with ideation and debugging. These efforts show the promise of automated support in Scratch, but they do not offer a standardized benchmark for model evaluation. More broadly, LLM-based program repair has been surveyed recently~\cite{yang2025surveyllmbasedautomatedprogram}, but existing results are confined to text code. In summary, while prior work has explored Scratch analysis/testing and repair techniques, \app is, to our knowledge, the first executable benchmark combining large-model evaluation with the unique demands of block-based programming.

\para{Analysis and Testing Tooling.}
A line of work has built static analysis~\cite{ConfigX,Clef} and automated testing~\cite{VeriCI} infrastructure to support assessment and quality assurance. Hairball provides lint-inspired static checks over Scratch projects~\cite{boe2013hairball}, and Dr.~Scratch automatically analyzes projects to assess computational thinking skills and flag common issues~\cite{moreno2015drscratch}. LitterBox detects recurring bug patterns and code smells in Scratch code~\cite{fraser2021litterbox}. On the dynamic side, Whisker enables automated and property-based testing for Scratch programs~\cite{stahlbauer2019whisker}, and has been extended with automated test generation and model-based testing for event-driven behaviors~\cite{whisker_modelbased2022}. These tools are valuable for educators and program analysis, but they do not provide an executable benchmark with paired buggy/fixed versions and a standardized repair protocol for large-model evaluation. Recent frameworks such as LitterBox+ explore integrating LLMs with Scratch static analysis to answer queries about projects~\cite{fein2025litterboxplus}, but they similarly focus on analysis rather than repair benchmarking.

\para{LLM Support for Scratch Programming.}
Recent work has explored large language models (LLMs) as copilots for Scratch, primarily to support creativity, ideation, and planning. Systems such as ChatScratch~\cite{chen2024chatscratch}, Cognimates Copilot, and MindScratch~\cite{chen2025mindscratch} embed LLMs into storyboarding or mind-mapping interfaces, helping learners generate assets and structure projects around learning goals. In parallel, LLMs have been applied to automated debugging and feedback in text-based programming, including bug detection, repair, and hint generation~\cite{legoues2015manybugs,deiner2023autotest,huang2025seeing,chen2021codex,zhang2025systematicstudytimelimit,PyDex,Gmerge}. However, LLM-based debugging remains unreliable, often producing hallucinated feedback, excessive edits, or full rewrites~\cite{song2023maps,Shao0S00025,Shi00025}, and these issues are amplified in Scratch, where correctness is visually grounded and behaviorally implicit.

\section{Conclusion}

We introduced \app, an executable benchmark for evaluating LLM understanding and repair on Scratch programs. \app contains 100 semantically complex projects, each paired with a single controlled, reversible bug injection, its inverse patch as the minimal gold fix, and a VM-executed test suite (interaction scenarios plus rerun-stabilized oracle assertions). This design enables closed-loop, execution-based evaluation of (i) functional repair success, (ii) patch minimality and semantic fidelity, and (iii) explanation faithfulness to trigger--mechanism--outcome semantics. Using \app, we presented the first systematic study of LLM behavior on block-based, event-driven debugging and repair, highlighting persistent challenges in grounding fixes and avoiding over-editing. Upon acceptance, we will publicly release \app and the evaluation harness to support comparisons and to accelerate future progress on LLMs for block-based programming.


\bibliographystyle{ACM-Reference-Format}
\bibliography{scratch}

\end{document}